\documentclass{tlp}

\usepackage{amssymb,latexsym,stmaryrd,epsfig,psfig}

\def\setmcodes#1#2#3{{\count0=#1 \count1=#3
    \loop \global\mathcode\count0=\count1 \ifnum \count0<#2
    \advance\count0 by1 \advance\count1 by1 \repeat}}

\def\defemb#1#2{\expandafter\def\csname #1\endcsname
  {\relax\ifmmode #2\else\hbox{$#2$}\fi}}

\defemb{cA}{{\cal A}}
\defemb{cB}{{\cal B}}
\defemb{cD}{{\cal D}}
\defemb{cE}{{\cal E}}
\defemb{cG}{{\cal G}}
\defemb{cI}{{\cal I}}
\defemb{cP}{{\cal P}}
\defemb{cN}{{\cal N}}
\defemb{cM}{{\cal M}}
\defemb{cS}{{\cal S}}
\defemb{cT}{{\cal T}}
\defemb{cV}{{\cal V}}

     {\end{tabular}\end{tt}\end{small}}

\def\anno#1{{\ooalign{\hfil\raise.07ex\hbox{\small{\rm #1}}\hfil%
        \crcr\mathhexbox20D}}}

\def\ll{[\![}
\def\rr{]\!]}
\def\Den#1{\relax\ifmmode \ll #1\rr \else\hbox{$\ll #1\rr$}\fi}

\def\posgd{\Den{P^G}_{pos}^{calls}}

\def\sizebin{\Den{P}_{size}^{bin}}
\def\dchk{\mathsf{{CHK}}}
\def\dinf{\mathsf{{INF}}}

\def\lwedge#1#2{\mathrel{\mathop{\wedge}\limits_{#1}^{#2}}}
\def\lbwedge#1#2{\mathrel{\mathop{\bigwedge}\limits_{#1}^{#2}}}
\def\lSigma#1#2{\mathrel{\mathop{\Sigma}\limits_{#1}^{#2}}}

\def\lbvee#1#2{\mathrel{\mathop{\bigvee}\limits_{#1}^{#2}}}

\newcommand{\set}[1]{\left\{
    \begin{array}{l}#1
    \end{array}
  \right\}}
\newcommand{\sset}[2]{\left\{~#1  \left|
                               \begin{array}{l}#2\end{array}
                          \right.     \right\}}

\newcommand{\Pos}{\textsf{Pos}}

\newtheorem{theorem}{Theorem}[section]
\newtheorem{example}{Example}[section]
\newtheorem{definition}{Definition}[section]

\submitted{24 July 2003}
\revised{}
\accepted{13 October 2003}

\title[Inferring Termination Conditions]
      {Inferring Termination Conditions for Logic  
        Programs using Backwards Analysis}

\author[SAMIR GENAIM and MICHAEL CODISH]
       {SAMIR GENAIM\\
         Dipartimento di Informatica\\
         Universit{\`a} degli Studi di Verona
       \and  MICHAEL CODISH\\
         Department of Computer Science\\
         Ben-Gurion University}

\begin{document}

\maketitle

\begin{abstract}
  This paper focuses on the inference of modes for which a logic program is
  guaranteed to terminate. This generalises traditional termination analysis
  where an analyser tries to verify termination for a specified mode.
  Our contribution is a methodology in which components of traditional
  termination analysis are combined with backwards analysis to obtain an
  analyser for termination inference.
  We identify a condition on the components of the analyser which guarantees
  that termination inference will infer all modes which can be checked to
  terminate.
  The application of this methodology to enhance a traditional termination
  analyser to perform also termination inference is demonstrated.
\end{abstract}

\section{Introduction}
\label{sec:intro}

This paper focuses on the inference of modes for which a logic program is
guaranteed to terminate.  This generalises traditional termination analysis
where an analyser tries to verify termination for a specified mode. For
example, for the classic $append/3$ relation, a standard analyser will
determine that a query of the form $append(x,y,z)$ with $x$ bound to a closed
list terminates and likewise for the query in which $z$ is bound to a closed
list. In contrast, termination inference provides the result
$append(x,y,z)\leftarrow x\lor z$ with the interpretation that the query
$append(x,y,z)$ terminates if $x$ or $z$ are bound to closed lists. We refer to
the first type of analysis as performing \emph{termination checking} and to the
second as \emph{termination inference}.
We consider universal termination using Prolog's leftmost selection rule and we
assume that unifications do not violate the occurs check.

Several analysers for termination checking are described in the literature. We
note the TermiLog system described in \cite{LinSag97} and the system based on
the binary clause semantics described in \cite{CT:JLP}.
Termination inference is considered previously by Mesnard and coauthors in
\cite{Mesnard:jicslp96,Mesnard:sas01,Mesnard:tocl}.  Here, we make the
observation that the missing link which relates termination checking and
termination inference is \emph{backwards analysis}. Backwards analysis is
concerned with the following type of question: Given a program and an assertion
at a given program point, what are the weakest requirements on the inputs to
the program which guarantee that the assertion will hold whenever execution
reaches that point.

In a recent paper, King and Lu \cite{king} describe a framework for backwards
analysis for logic programs in the context of abstract interpretation. In their
approach, the underlying abstract domain is required to be condensing or
equivalently, a complete Heyting algebra. This property ensures the existence
of a weakest requirement on calls to the program which guarantees that the
assertions will hold.

To demonstrate this link between termination checking and termination
inference, we apply the framework for backwards analysis described by King and
Lu \cite{king} to enhance the termination (checking) analyser described in
\cite{CT:JLP} to perform also termination inference. We use the condensing
domain $\Pos{}$, of positive Boolean formula, to express the conditions on the
instantiation of arguments which guarantee the termination of the program.

The use of a standard framework for backwards analysis provides a formal
justification for termination inference and leads to a simple and efficient
implementation similar in power to that described in \cite{Mesnard:sas01}. It
also facilitates a formal comparison of termination checking and inference. In
particular, we provide a condition on the components of the analyser which
guarantee that termination inference will infer all modes which termination
checking can prove to be terminating.

In the rest of the paper, Section \ref{prelim} provides some background and a
motivating example. Section \ref{sec:b-analysis} reviews the idea of backwards
analysis.  Section \ref{sec:from-chk-to-inf} illustrates how to combine
termination analysis with backwards analysis in order to obtain termination
inference and investigates their relative precision. Section \ref{sec:exper}
presents an experimental evaluation.  Finally, Section \ref{sec:rw} reviews
related work and Section \ref{sec:conc} concludes. A preliminary version of
this paper appeared as Ref.  \cite{GC:LPAR01}.  Our implementation
\cite{terminweb} can be accessed on the web. It supports termination checking
as described in \cite{CT:JLP} and termination inference as described in this
paper.

\section{Preliminaries and Motivating Example}
\label{prelim}

We assume a familiarity with the standard definitions for logic programs
\cite{Lloyd87,Apt88} as well as with the basics of abstract interpretation
\cite{Cou-Cou:POPL77,Cou-Cou:JLP92}.  This section describes the standard
program analyses upon which we build in the rest of the paper. For notation, in
brief: variables in logic programs are denoted as in Prolog (using the upper
case) while in relations, Boolean formula, and other mathematical context we
use the lower case. We let $\bar x$ denote a tuple of distinct variables
$x_1,\ldots,x_n$. To highlight a specific point in a program we use labels of
the form $\anno{a}$.

Size relations and instantiation dependencies rest at the heart of termination
analysis: size information to infer that some measure on program states
decreases as computation progresses; and instantiation information, to infer
that the underlying domain is well founded. Consider the recursive clause of
the $append/3$ relation: $append([X|Xs],Ys,[X|Zs]) \leftarrow
append(Xs,Ys,Zs).$
It does not suffice to observe that the size of the first and third arguments
decrease in the recursive call. To guarantee termination one must also ensure
that at least one of these arguments is sufficiently instantiated in order to
argue that this recursion can be activated only a finite number of times.

Instantiation information is traditionally obtained through abstract
interpretation over the domain \Pos\ which consists of the positive Boolean
functions augmented with a bottom element (representing the formula $false$).
The elements of the domain are ordered by implication and represent equivalence
classes of propositional formula.
This domain is usually associated with its application to infer groundness
dependencies where a formula of the form $x \land (y \rightarrow z)$ is
interpreted to describe a program state in which $x$ is definitely bound to a
ground term and there exists an instantiation dependency such that whenever $y$
becomes bound to a ground term then so does $z$.  Similar analyses can be
applied to infer dependencies with respect to other notions of instantiation.
Boolean functions are used to describe the groundness dependencies in the
success set of a program $P$ as well as in the set of calls which arise in the
computations for an initial call pattern $G$.  We denote these approximations
by $\Den{P}_{pos}^{suc}$ and $\posgd$ respectively. The elements are of the
form $p(\bar x) \leftarrow \varphi$ where $p/n$ is a predicate defined in $P$
and $\varphi$ is a positive Boolean function on $\bar x$.
For details on \Pos\ see \cite{Mar-Son:loplas93}.

Size relations express linear information about the sizes of terms (with
respect to a given norm function) \cite{DeSchreye/Verschaetse:95,Ka76}. For
example, the relation $x\leq z ~\wedge~ y\leq z$ describes a program state in
which the sizes of the terms associated with $x$ and $y$ are less or equal to
the size of the term associated with $z$. Similarly, a relation of the form
$z=x+y$ describes a state in which the sum of the sizes of the terms associated
with $x$ and $y$ is equal to the size of the term associated with $z$.
Here the variables represent sizes and hence are implicitly constrained to be
non-negative.
Several methods for inferring size relations are described in the literature
\cite{BenoyKing,BroSag89,cousothalb:78,DeSchreye/Verschaetse:95}. They differ
primarily in their approach to obtaining a finite analysis as the abstract
domain of size relations contains infinite chains.
For a survey on termination analysis of logic programs see
\cite{DeSchreye:JLP94}.

Throughout this paper we will use the so-called term-size norm for size
relations for which the corresponding notion of instantiation is groundness.
We base our presentation on the termination (checking) analyser described in
\cite{CT:JLP} although we could use as well almost any of the alternatives
described in the literature.  This analyser is based on a bottom-up $T_P$ like
semantics which makes loops observable in the form of binary clauses. This
provides a convenient starting point for termination inference as derived in
this paper.  We denote the abstraction of this semantics for a program $P$ over
the domain of size relations as $\sizebin$.  Each element of $\sizebin$
represents a loop and is of the form $p(\bar x) \leftarrow \pi, p(\bar y)$
where $\pi$ is a conjunction of linear constraints. In the examples these are
represented as lists of constraints.

We proceed to demonstrate our approach by example in four steps:

~\\\noindent\textsf{\textbf{The first step:}}
Consider the $append/3$ relation. 
\begin{quote}\tt
\begin{tabbing}
 xxxx \= xxx \= xxxxxxxxxxxxxxxxx \=\kill
\>     append([X|Xs],Ys,[X|Zs]) :- append(Xs,Ys,Zs). \\
\>     append([],Ys,Ys).                             
\end{tabbing}
\end{quote}
\noindent
Termination checking  reports a single abstract binary
clause:
\begin{quote}\tt
\begin{tabbing}
 xxxx \= xxx \= xxxxxxxxxxxxxxxxx \=\kill
\>   append(A,B,C) :- [D<A, F<C, B=E], append(D,E,F).
\end{tabbing}
\end{quote}
indicating that subsequent calls $append(A,B,C)$ and $append(D,E,F)$ in a
computation, involve a decrease in size for the first and third arguments
($D<A$ and $F<C$) and maintain the size of the second argument ($B=E$).
To guarantee that this loop may be traversed only a finite number of times, it
is sufficient to require that either $A$ or $C$ be sufficiently instantiated.
This can be expressed as a Boolean condition: $append(x,y,z) \leftarrow (x\lor
z)$.

Backwards analysis is now applied to infer the weakest conditions on the
program's predicates which guarantee this condition. For this example the
inference is complete and we have derived the result: $append(x,y,z) \leftarrow
x\lor z$ interpreted as specifying that $append(x,y,z)$ terminates if $x$ or
$z$ are bound to ground terms.

~\\\noindent\textsf{\textbf{The second step:}}
Consider the use of $append/3$ to define list membership. Adding the following
clause to the program introduces no additional loops:
\begin{quote} \tt
\begin{tabbing}
 xxxx \= xxx \= xxxxxxxxxxxxxxxxx \=\kill
\>     member(X,Xs) :- append(A,[X|B],Xs).       
\end{tabbing}
\end{quote}
Backwards analysis should specify the weakest condition on $member(X,Xs)$ which
guarantees the termination condition $A\lor Xs$ for $append(A,[X|B],Xs)$. This
is obtained through projection which for backwards analysis is defined in terms
of universal quantification as $\forall_A.(A\lor Xs)$.
The resulting Boolean precondition is: $member(x,y)\leftarrow y$ indicating
that $member(x,y)$ terminates if $y$ is ground.

~\\\noindent\textsf{\textbf{The third step:} }
We now add to the program a definition for the subset/2 relation:
\begin{quote} \tt
\begin{tabbing}
 xxxx \= xxx \= xxxxxxxxxxxxxxxxx \=\kill
\>     subset([X|Xs],Ys) :-  member(X,Ys), subset(Xs,Ys). \\
\>     subset([],Ys).                
\end{tabbing}
\end{quote}
Termination checking reports an additional loop:
\begin{quote} \tt
\begin{tabbing}
 xxxx \= xxx \= xxxxxxxxxxxxxxxxx \=\kill
\>       subset(A,B) :- [B=D,C<A], subset(C,D).
\end{tabbing}
\end{quote}
which will be traversed a finite number of times if $A$ is sufficiently
instantiated. For the first clause to terminate both loops must terminate: for
$append/3$ in the call to $member(X,Ys)$ and for $subset/2$ in the call to
$subset(Xs,Ys)$. So both $Xs$ and $Ys$ must be instantiated which implies that
both arguments of $subset/2$ should be ground inputs.  Namely, $subset(x,y)
\leftarrow x\wedge y$.

~\\\noindent\textsf{\textbf{The fourth step:}}
This step demonstrates that the precondition on a call in a clause body may be
(partially) satisfied by answers to calls which precede it. Consider adding to
the program a clause:
\begin{quote} \tt
\begin{tabbing}
 xxxx \= xxx \= xxxxxxxxxxxxxxxxx \=\kill
\>    s(X,Y,Z) :- \anno{a} append(X,Y,T), \anno{b} subset(T,Z).
\end{tabbing}
\end{quote}
which defines a relation $s(x,y,z)$ such that the set $z$ contains the union of
sets $x$ and $y$. The preconditions for termination derived in the previous
steps specify the conditions $x\lor t$ and $t\land z$ at points \anno{a} and
\anno{b} respectively. In addition, from a standard groundness analysis we know
that on success $append(x,y,t)$ satisfies $(x\land y) \leftrightarrow t$. So,
instead of imposing on the clause head both conditions from the calls in its
body, as we did in the previous step, we may weaken the second condition in
view of the results from the first call. Namely $(((x\land y) \leftrightarrow
t) \rightarrow t\land z)$.
Now the termination condition inferred for $s(x,y,z)$ is~~ $\forall_t.((x\lor
t) \land (((x\land y) \leftrightarrow t) \rightarrow t\land z)) \equiv x \land
y\land z.  $

In general, the steps illustrated above, though sufficient for these simple
examples, do need to be applied in iteration. In the next section we describe
more formally the steps required for backwards analysis.

\section{Backward Analysis}
\label{sec:b-analysis}

This section presents an abstract interpretation for backwards analysis using
the domain \Pos\ distilled from the general presentation given in \cite{king}.
Clauses are assumed to be normalised and contain assertions so that they are of
the form $h(\bar{x}) \leftarrow \mu \diamond b_1,\ldots,b_n$
where $\mu$ is a \Pos\ formula, interpreted as an instantiation condition that
must hold when the clause is invoked, and $b_i$ is either an atom, or a
unification operation.

The analysis associates preconditions, specified in \Pos, with the predicates
of the program. Initialised to $true$ (the top element in $Pos$) these
preconditions become more restrictive (move down in \Pos) through iteration
until they stabilise.
At each iteration, clauses are processed from right to left using the current
approximations for preconditions on the calls together with the results of a
standard groundness analysis to infer new approximations for these
preconditions.

For the basic step, consider a clause of the form: $p \leftarrow \ldots~
\anno{a}, q, \anno{b} ~\ldots$
and assume that the current approximation for the precondition for a predicate
$q$ is $\varphi_q$, the success of $q$ is approximated by $\psi_q$, and that
processing the clause from right to left has already propagated a condition
$e_b$ at the point \anno{b}. Then, to insure that $e_b$ will hold after the
success of $q$, it suffices to require at \anno{a} the conjunction of
$\psi_{q}$ with the weakest condition $\sigma$ such that $(\sigma\wedge\psi_q)
\rightarrow e_b$.  This $\sigma$ is precisely the pseudo-complement \cite{GS98}
of $\psi_q$ with respect to $e_b$, obtained as $\psi_q\rightarrow e_b$.  So
propagating one step to the left gives the condition $e_a = \varphi_q \wedge
(\psi_q\rightarrow e_b)$.

Now consider a clause $h(\bar x) \leftarrow \mu \diamond b_1,\ldots,b_n$ with
an assertion $\mu\in\Pos$.
Assume that the current approximation for the precondition of $h(\bar x)$ is
$\varphi$ and let $\psi_i$ and $\varphi_i$ denote respectively the
approximation of the success set of $b_i$ (obtained through standard groundness
analysis) and the current precondition for $b_i$ ($1\le i\le n$).
Backwards analysis infers a new approximation $\varphi'$ of the precondition
for $h(\bar x)$ by consecutive application of the basic step described above.
We start with $e_{n+1}=true$ and through $n$ steps (with $i$ going from $n$ to
$1$) compute a condition $e_i=\varphi_i \wedge (\psi_i \rightarrow e_{i+1})$
which should hold just before the call to $b_i$. After computing $e_1$ we take
$e_0=\mu\wedge e_1$ and project $e_0$ on the variables $\bar x$ of the head by
means of universal quantification. The new condition is finally obtained
through conjunction with the previous condition $\varphi$. Namely,
$\varphi'=\varphi\wedge \bar\forall\bar x.~e_0$.

There is one subtlety in that $\Pos$ is not closed under universal
quantification. To be precise, elimination of $x$ from $\sigma\in\Pos$ is
defined as the largest element in \Pos\ which implies $\forall_x.\sigma$. When
$\forall_x.\sigma$ is not positive then the projection gives $false$ which is
the bottom element in \Pos.

\begin{example}
  Consider the clause 
\begin{quote} \tt
\begin{tabbing}
 xxxx \= xxx \= xxxxxxxxxxxxxxxxx \=\kill
\>    subset(A,B) :-
      \anno{{$e_0$}} A $\diamond$
        \anno{{$e_1$}} A=[X|Xs],
        \anno{{$e_2$}} B=Ys, \\
\>\>      \anno{{$e_3$}} member(X,Ys),
        \anno{{$e_4$}} subset(Xs,Ys) \anno{$e_5$}.
\end{tabbing}
\end{quote}
where the assertion $A$ states that the first argument must be ground and the
success patterns (derived by a standard groundness analysis) and the current
approximation of the preconditions are
(respectively):\\
\begin{minipage}{6.5cm}
\[
\Psi = \left\{
\begin{array}{l}
member(x,y) \leftarrow (y \rightarrow x)\\
subset(x,y) \leftarrow (y \rightarrow x)
\end{array}
\right\}
\]
\end{minipage}
\begin{minipage}{6cm}
\[
\Phi = \left\{
\begin{array}{l}
 member(x,y) \leftarrow y\\
 subset(x,y) \leftarrow x\\
\end{array}
\right\}.
\]
\end{minipage}
~\\Starting from $e_5=true$, the conditions $e_4,\ldots,e_1$ are obtained by
substituting in $e_i=\varphi_i\wedge(\psi_i\rightarrow e_{i+1})$ as illustrated
in the following table:

\[\begin{array}{c||l|l||c}
  i & \varphi_i & \psi_i &
                    e_i = \varphi_i\wedge(\psi_i\rightarrow e_{i+1})\\
  \hline
  4 & Xs & Ys\rightarrow Xs & true  \\
  3 & Ys & X\rightarrow Xs  & Ys \wedge (X\rightarrow Xs) \\
  2 & true & B\leftrightarrow Ys &
     (B\leftrightarrow Ys)\rightarrow (Ys\wedge (X\rightarrow Xs))\\
  1 & true & A\leftrightarrow (X\wedge Xs) &
     (A\leftrightarrow (X\wedge Xs)) \rightarrow 
     (B\leftrightarrow Ys)\rightarrow (Ys\wedge (X\rightarrow Xs))\\
\end{array}
\]
We now obtain $e_0$ as $A\wedge e_1$ and projecting $e_0$ to the variables in
the head gives $\forall_{Xs,Ys,X}.(e_0) = A \wedge B$.  Which leads to the new
precondition $subset(x,y) \leftarrow x\land y$.

\end{example}

In \cite{king}, the authors formalise backwards analysis as the greatest fixed
point of an operator over \Pos. In our implementation \cite{terminweb}
backwards analysis is realised as a simple Prolog interpreter which manipulates
Boolean formula using a package for binary decision diagrams written by
Armstrong and Schachte (used in \cite{Arm-Mar-Sch-Son:SCP98} and described in
\cite{Schachte:phd}).

\section{From termination checking to termination Inference}
\label{sec:from-chk-to-inf}

Termination checking aims to determine if a program is guaranteed to terminate
for a specified mode. Termination inference aims to infer a set of modes for
which the program is guaranteed to terminate. To be precise, we introduce the
following definition and terminology.
\begin{definition}[Mode]
  A mode is a tuple of the form $p(m_1, \ldots ,m_n)$ where $m_i$ $(1\le i\le
  n)$ is either \textbf{b} (``bound'') or \textbf{f} (``free''). We can view a
  mode as a call pattern $p(\bar x) \leftarrow \varphi$ where $\varphi =
  \wedge\sset{x_i}{x_i=\mathbf{b}\wedge(1\le i\le n)}$.
\end{definition}
Given a norm function, we say that a program terminates for a mode $p(m_1,
\ldots ,m_n)$ if it terminates for all initial queries $p(t_1,\ldots,t_n)$ such
that for $1\le i\le n$,
$m_i=\mathbf{b}$ implies that $t_i$ is rigid with respect to the given norm.

This section describes how an analyser for termination inference can be derived
from an analyser for termination checking together with a component for
backwards analysis.  We first describe in Section \ref{sec:chk} the activities
performed by an analyser for termination checking. Then, in Section
\ref{sec:inf} we explain how some of these activities are combined with a
backwards analysis component to obtain an analyser for termination inference.
Finally, in Section \ref{cvsi}, we compare the precision of termination
checking and inference.

\subsection{Termination Checking}
\label{sec:chk}

Termination checking involves two activities: first, the loops in the program
are identified and characterised with respect to size information; and second,
given the mode of an initial query, it is determined if for each call pattern
in a computation and for each loop, some measure on the sizes of some of the
sufficiently instantiated arguments in the call decrease as the loop
progresses.

In the analyser described in \cite{CT:JLP} these activities are performed in
two phases.
The first (goal independent) phase computes a set of abstract binary clauses
$\sizebin$ which describe, in terms of size information, the loops in the
program $P$.
The second (goal dependent) phase determines a set of call patterns $\posgd$
for a initial mode $G$ and checks that for each call in $\posgd$ and each
corresponding loop in $\sizebin$ there exists a suitable well-founded
decreasing measure.
The next definitions provide the notions required to state the theorem which
follows (reformulating Proposition 6.5 in \cite{CT:JLP}) to provide a
sufficient termination (checking) condition.

\begin{definition}[Decreasing arguments set] 
 \label{d.a.s.}
 A set of arguments $I=\{x_{i_1},\ldots,x_{i_k}\} \subseteq \bar x$ is
 \textsf{decreasing} for an abstract binary clause $\beta=p(\bar x) \leftarrow
 \pi,p(\bar y)$ if there exist coefficients $a_1,\ldots,a_k$ such that $\pi
 \models a_1x_{i_1}+\cdots+a_kx_{i_k} > a_1y_{i_1}+\cdots+a_ky_{i_k}$. The set
 of all decreasing sets of arguments for $\beta$ is denoted by $\cD(\beta)$.
\end{definition}

Note that by definition $\cD(\beta)$ is closed under extension. Namely, if
$I\in\cD(\beta)$ and $I'\supseteq I$ then $I'\in\cD(\beta)$ (simply map
coefficients for the arguments in $I' \setminus I$ to 0).

\begin{definition}[Instantiated arguments set] 
\label{def:ie}
We say that a set of arguments $I \subseteq \bar{x}$ is instantiated in a call
pattern $\kappa = p(\bar x) \leftarrow \varphi$ if $\varphi \models
\wedge\sset{x}{x\in I}$. We denote by $I_\varphi$ the set of all arguments
instantiated in $\kappa$.
\end{definition}

\begin{theorem}[Termination Condition]
\label{tc}
Let $P$ be a logic program and $G$ an initial call pattern. If for each call
pattern $\kappa= p(\bar x) \leftarrow \varphi \in \posgd$ and corresponding
binary clause $\beta= p(\bar x) \leftarrow \pi,p(\bar y) \in \sizebin$ there
exists a set of arguments $I\subseteq\bar x$ which is \textsf{instantiated} in
$\kappa$ and \textsf{decreasing} for $\beta$ then $P$ terminates for $G$.
\end{theorem}

\begin{example}
  The analysis of the $append/3$ relation (detailed in Section
  \ref{prelim}) for the initial mode $G\equiv append(b,b,f)$ gives:
  \[\begin{array}{l}
     \sizebin=
        \set{\mathtt{append(A,B,C) 
                 \leftarrow [D<A, F<C, B=E], ~append(D,E,F)}}\\[1ex]
      \posgd =\set{\mathtt{append(A,B,C) \leftarrow A\wedge B}}
  \end{array}\]
  The termination condition holds for this single binary clause and call
  pattern with $I=\{A\}$ as well as with $I=\{A,B\}$.
\end{example}

We now focus in on that component of the termination checker that checks if the
termination condition is satisfied for a call pattern $p(\bar x)
\leftarrow\varphi$ and a corresponding binary clause $\beta$.
We denote by $\dchk(I,\beta)$ the decision procedure which is at the heart of
this component and determines if some subset of $I$ is decreasing for $\beta$.
Since any decreasing and instantiated enough set of arguments is a subset of
$I_\varphi$, the analyser will typically invoke $\dchk(I_\varphi,\beta)$.

For the correctness of termination checking, $\dchk(I,\beta)$ must be sound but
need not be complete. Namely if $\dchk(I,\beta)$ reports \textsf{``yes''} then
$I$ must be a decreasing set of arguments for $\beta$.
The termination analyser described in \cite{CT:JLP} applies a simple (and fast)
decision procedure which is not complete but works well in practise.
For a call $p(\bar x) \leftarrow \varphi$ with instantiated variables
$I_{\varphi}=\{x_{i_1},\ldots,x_{i_k}\}$ and a matching binary clause $p(\bar
x) \leftarrow \pi, p(\bar y)$ the system checks if $\pi \models
y_{i_1}+\cdots+y_{i_k} \ge x_{i_1}+\cdots+x_{i_k}$ (recall that all of the
variables are non-negative).
If not, then it reports ``yes'' because it must be the case that for some $1\le
j\le k$, $y_{i_j}<x_{i_j}$ and hence the singleton $\{x_{i_j}\}$ is a
decreasing argument set.

A complete procedure for $\dchk$ (denoted \textsf{SVG}) is described in
\cite{SVG} and discussed also in \cite{Mesnard:sas01}. There the authors
observe that checking the satisfiability of the \emph{non-linear} constraint
system $\pi \wedge \exists a_1,\ldots,a_k.(a_1x_{i_1} + \cdots + a_kx_{i_k} >
a_1y_{i_1} + \cdots + a_ky_{i_k})$,
for coefficients $a_1,\ldots,a_k$, is equivalent to checking that of the dual
constraint system which is linear. See the references above for details. The
TerminWeb analyser \cite{terminweb} offers the optional use of this procedure.

\subsection{Termination Inference}
\label{sec:inf}

Our approach to termination inference proceeds as follows:
(1) The first phase of the termination checker is applied to approximate the
loops in the program as binary clauses with size information ($\sizebin$);
(2) Each loop in $\sizebin$ is examined to extract an initial (Boolean)
termination assertion on the instantiation of arguments of the corresponding
predicate which guarantee that the loop can be executed only a finite number of
times; and
(3) Backwards analysis is applied to infer the weakest constraints on the
instantiation of the initial queries to guarantee that these assertions will be
satisfied by all calls.

Intuitively, an initial termination assertion for a predicate $p(\bar x)$ is a
Boolean formula constructed so as to guarantee that each binary clause has at
least one set of arguments which is instantiated enough and decreasing.
To this end, the best we can do for a given binary clause $\beta$ is to require
the instantiation of the variables in (at least) one of of the decreasing sets
of arguments in $\cD(\beta)$ (a disjunction). This gives the most general
initial termination assertion for $\beta$. For a predicate in the program, the
assertions for all of its binary clauses must hold (a conjunction).
In practise, an analyser for termination inference involves a component
$\dinf(\beta)$ which approximates $\cD(\beta)$ (from below) for an abstract
binary clause $\beta$. For the correctness of termination inference,
$\dinf(\beta)$ must be sound but need not be complete. Namely it may return a
subset of $\cD(\beta)$. Of course if it is complete (i.e. computes
$\cD(\beta)$) then the inference will be more precise.
Given such a procedure $\dinf(\beta)$, the initial termination assertions are
specified as follows:
\begin{definition}[Initial Termination Assertion]
\label{def:assertion}
Let $P$ be a logic program.
The initial termination assertions for a binary clause $\beta \in \sizebin$,
and a predicate $p/n \in P$ are given as:
  \[
    \mu(\beta) = \lbvee{I\in\dinf(\beta)}{}\!\!\!
                   \left(\lwedge{x\in I}{}\!\!x\right)
  \hspace{2cm} 
  \mu(p(\bar{x})) = \lbwedge{\beta\in B}{}\mu(\beta)
   \] 
  where $B\subseteq \sizebin$ is the set of binary clauses for
  $p(\bar x)$ in $\sizebin$.
\end{definition}

Note that we can assume without loss of generality that $\dinf$ is closed under
extension as the assertions $\mu(\beta)$ are invariant to the addition of
extending sets of arguments.

\begin{example}
\label{ex:merge}
  Consider as $P$ the $split/3$ relation (from merge sort):
\begin{quote} \tt
\begin{tabbing}
 xxxx \= xxx \= xxxxxxxxxxxxxxxxx \=\kill
\> split([],[],[]).\\
\> split([X|Xs],[X|Ys],Zs) :- split(Xs,Zs,Ys).
\end{tabbing}
\end{quote}
The binary clauses obtained by the analyser of \cite{CT:JLP} are:
\begin{itemize}
\item[] $\beta_1 =~~ split(x_1,x_2,x_3) \leftarrow 
    [y_1<x_1,y_3<x_2,x_3=y_2], ~split(y_1,y_2,y_3).$
\item[] $\beta_2 =~~ split(x_1,x_2,x_3) \leftarrow 
    [y_1<x_1,y_2<x_2,y_3<x_3], ~split(y_1,y_2,y_3).$
\item[] $\beta_3 =~~ split(x_1,x_2,x_3) \leftarrow 
    [y_1<x_1,y_3<x_2,y_2<x_3], ~split(y_1,y_2,y_3).$
\end{itemize}
%m
Here, $\beta_1$ represents the size information corresponding to passing one
time through the loop defined by the second clause; $\beta_2$ the information
corresponding to any even number of times through the loop; and $\beta_3$ any
odd number of times (greater than 1).

Let $S\!\!\uparrow$ denote the closure of a set $S$ under extension with
respect to the variables of interest.
Assuming that
$\dinf(\beta_1)=\dinf(\beta_3)=\{\{x_1\},\{x_2,x_3\}\}\!\!\!\uparrow$ (note
that $y_1<x_1$ and $y_2+y_3<x_2+x_3$) and $\dinf(\beta_2)=\{\{x_1\}, \{x_2\},
\{x_3\}\}\!\!\!\uparrow$ (note that $y_1<x_1,y_2<x_2,y_3<x_3$),
we have $\mu(\beta_1) = \mu(\beta_3) = x_1 \vee (x_2 \wedge x_3)$; and
$\mu(\beta_2) = x_1 \vee x_2 \vee x_3$.
The assertion for $split/3$ is:
$  \mu(split(x_1,x_2,x_3)) =
     \mu(\beta_1)\wedge\mu(\beta_2)\wedge\mu(\beta_3) =
     x_1 \vee (x_2 \wedge x_3).
$
Backwards analysis starting from this assertion infers the termination
condition $x_1 \vee (x_2 \wedge x_3)$ for $split(x_1,x_2,x_3)$.
\end{example}

The result of backwards analysis is a positive Boolean formula for each
predicate describing the conditions under which a corresponding initial query
terminates. The following definition specifies how the initial modes for
terminating queries are derived from this formula.

\begin{definition}[Terminating mode]
  Let $P$ be a logic program. We say that $p(m_1, \ldots ,m_n)$ is terminating
  for $p(x_1,\ldots,x_n)$ defined in $P$ if the conjunction $\wedge\{x_i
  ~|~m_i=b\}$ implies the condition inferred by termination inference for
  $p(\bar x)$.
\end{definition}

\begin{example}
  Consider again the $split/3$ relation given in Example \ref{ex:merge} for
  which we inferred $\mu(split(x_1,x_2,x_3)) =x_1 \vee (x_2 \wedge x_3)$.  Both
  $split(b,f,f)$ and $split(f,b,b)$ are terminating modes because $x_1$ and
  $(x_2 \wedge x_3)$ imply $x_1 \vee (x_2 \wedge x_3)$.
\end{example}

The correctness of the method described follows from the results of
\cite{CT:JLP} and \cite{king}.

\begin{theorem}\label{th:2}
  Let \/ $\dinf$ be a sound procedure, $P$ a logic program and $p(\bar m)$ a
  terminating mode for $p(\bar x)$ inferred by termination inference. Then $P$
  terminates for $p(\bar m)$.
\begin{proof}
\noindent
Let $G=p(\bar t)$ be an initial query described by the inferred terminating
mode $p(\bar m)$. The correctness of backwards analysis garantees that when
executing $G$, any call to a predicate $q/n$ satisfies the assertions inferred
for $q/n$.
From the specification of the initial termination assertion
(Definition~\ref{def:assertion}) we know that $\mu(q(\bar x)) \models
\mu(\beta)$ for each $\beta=q(\bar x) \leftarrow \pi, q(\bar y) \in \sizebin$.
Hence, at least one set of arguments for $\beta$ is decreasing and sufficiently
instantiated. This means that the termination condition of Theorem \ref{tc}
holds.
\end{proof}
\end{theorem}

In the analyzer for termination inference implemented in the context of this
work \cite{terminweb} we adopt for $\dinf$ a fast though incomplete procedure.
Given a binary clause $\beta=p(\bar x) \leftarrow \pi,p(\bar y)$ the procedure
works as follows where we denote the arguments of $p(\bar x)$ as
$\cI=\{1,\ldots,n\}$:
First, it computes the set $\cI'=\{ i ~|~ \pi \models x_i>y_i\}$ which includes
all argument positions that are decreasing. Each singleton subset of $\cI'$ is
reported by the procedure to be a decreasing set of arguments;
Second, it checks if the sum of the non-decreasing arguments is decreasing.
Namely, if
\[
\pi \models \lSigma{i\in\cI\setminus\cI'}{}x_i > 
                           \lSigma{i\in\cI\setminus\cI'}{}\!\!\!y_i
\]
If so, then it reports that $\cI\setminus\cI'$ is a decreasing set of
arguments.

Performing step 2 does appear to make a difference. This simplistic approach
works well in practice for the standard benchmarks and guarantees scalability
of the analysis.
For example consider the binary clause $\beta_1$ from Example \ref{ex:merge}.
The only decreasing singleton is $\{x_1\}$ and the set of all non-decreasing
arguments $\set{x_2,x_3}$ is also decreasing, this enables the detection of the
terminating mode $split(x_1,x_2,x_3) \leftarrow x_2\wedge x_3$.

In \cite{Mesnard:sas01}, the authors adopt a complete algorithm for $\dinf$
which they call \textsf{Extended SVG}.
Similar to \textsf{SVG} the authors consider the dual (linear) constraint
system of the form $\pi \wedge (a_1 x_{k_1} + \cdots + a_k x_{k_i} > a_1
y_{k_1} + \cdots + a_k y_{k_i})$. But instead of checking for satisfiability,
they look for the smallest subsets $\set{x_{k_1}, \cdots x_{k_i}} \subseteq\bar
x$ for which the constraint system is satisfiable. This is done by projecting
the system $\pi \wedge (a_1 x_{1} + \cdots + a_n x_{n} > a_1 y_{1} + \cdots +
a_n y_{n})$ on the variables $a_1,\ldots,a_n$ and systematically trying to bind
some of the $a_i$'s to zero.
In general this can require an exponential number of steps. However, the
author's experimentation indicates that the algorithm works well in practise.
See the reference above for details.

\subsection{Precision of Termination Checking vs. Inference}
\label{cvsi}

To compare the precision of an analyser for termination checking with one for
termination inference the relevant question is: Is there some mode which can be
checked to be terminating which is not inferred to be terminating (or vice
versa)?
In particular we would like to compare the precision of our own two analysers
for checking and inferring termination as well as with the cTI analyser for
termination inference.
In the next section we provide an experimental comparison for both efficiency
and precision. Here we are concerned with a theoretical comparison.

To keep all else the same, we will assume that the analysers being compared
obtain the same approximations of a program's loops ($\sizebin$ in our
terminology) and use the $Pos$ domain to approximate instantiation information.
For our two analysers these assumptions are of course true as we use the same
component to compute $\sizebin$.

Given that all other parameters in the analysers are the same, it is the
relation between the precision of the
specific choices for the
procedures $\dchk$ and $\dinf$ which determine the relevant precision of
termination checking and termination inference.
The comparison for a given choice of $\dchk$ and $\dinf$ is done by
considering for each abstract binary clause $\beta$ the sets $\dinf(\beta)$ and
$\sset{I}{\dchk(I,\beta)=``yes"}$.  If these sets are equal for all $\beta$
then we say that $\dchk$ and $\dinf$ are of the same accuracy.  In particular
if both $\dchk$ and $\dinf$ are complete then they are of the same accuracy,
As we have already noted, cTI employs an $\dinf$ procedure which is complete
and TerminWeb applies $\dchk$ and $\dinf$ procedures which are sound but not
complete.

The following theorem states that if $\dchk$ and $\dinf$ are of the same
accuracy then termination checking and inference report equivalent results.

\begin{theorem} Let $\cA_{tc}$ and $\cA_{ti}$ be analysers for 
  checking and inferring termination based on procedures $\dchk$ and $\dinf$ of
  the same accuracy and assume that these analysers approximate loops and
  instantiation information in the same way.  Assume also that $\cA_{ti}$ is
  based on backwards analysis.  Then, $\cA_{tc}$ reports that $P$ terminates
  for a mode $p(\bar m)$ if and only if $p(\bar m)$ is inferred by $\cA_{ti}$.
\begin{proof}
  Let us first make two simple observations concerning backwards analysis:
\begin{itemize}
\item[-]\textbf{(BA$_1$)}: Let $P$ be a logic program, $G=q(\bar m)$ an initial
  call pattern, $\psi_q=\wedge\sset{x_i}{m_i=b}$ and $P'$ a logic program with
  assertions defined by introducing to the clauses in $P$ the call patterns
  from $\posgd$ as initial assertions:
\[ P' = \sset{h(\bar x)\leftarrow \varphi\diamond body}
             {h(\bar x)\leftarrow body\in P,\\
              h(\bar x)\leftarrow \varphi\in\posgd}.
\]
Then, if $q(\bar x) \leftarrow \varphi_{q}$ is the result of backwards
analysis of $P'$ for $q(\bar x)$, then $\psi_{q} \models \varphi_{q}$.\\[1ex]

\item[-] \textbf{(BA$_2$)}: Let $P_1$ be a logic program with assertions and
  let $q(\bar x)\leftarrow\varphi_1$ be the result of backwards analysis of
  $P_1$ for $q/n$.
  Let $P_2$ be a program obtained by replacing an assertion $\mu_1$ in $P_1$ by
  an assertion $\mu_2$ such that $\mu_1\models\mu_2$ and let $q(\bar
  x)\leftarrow\varphi_2$ be the result of backwards analysis of $P_2$ for
  $q/n$.
Then $\varphi_1\models\varphi_2$.\\[1ex]
\end{itemize}
\noindent\textbf{$\Rightarrow$}~
Let $G=q(\bar m)$ be a mode for which $\cA_{tc}$ proves termination, we show
that $G$ is inferred by $\cA_{ti}$. Denote $\psi_q=\wedge\sset{x_i}{m_i=b}$ and
let $p(\bar x) \leftarrow \varphi \in \posgd$ and $\beta\equiv p(\bar x)
\leftarrow \pi,p(\bar y) \in \sizebin$. Consider the set $I_{\varphi}$ of
variables instantiated in $\varphi$.
$\dchk(I_{\varphi},\beta)$ answers ``yes'' because $\cA_{tc}$ proves
termination and by the assumption that $\dchk$ and $\dinf$ are of the same
accuracy, $I_{\varphi}\in\dinf(\beta)$. Hence, by
Definition~\ref{def:assertion}, $\wedge I_\varphi \models \mu(p(\bar x))$. By
Definition \ref{def:ie} $\varphi\models \wedge I_\varphi$, so we have
  $\varphi\models\mu(p(\bar x))$ \textbf{(*)}. 
  Let $q(\bar x) \leftarrow\varphi_q$ be the result of backwards analysis for
  $P$ with call patterns from $\posgd$ as initial assertions. By observation
  {\small\textbf{(BA$_1$)}} $q(\bar x) \leftarrow\varphi_q$ is the call pattern
  from $q/n$ and hence $\psi_q\models\varphi_q$ (because $G$ is one of the call
  patterns for $q/n$).
  
  Now by \textbf{(*)}, the termination assertions ($\mu(p(\bar x))$) are more
  general than the call patterns ($\varphi$) and hence by observation
  {\small\textbf{(BA$_2$)}} $\psi_q$ implies the result of backwards analysis
  with termination assertions replacing call patterns. In particular this is
  the case for $q(\bar x)$ and so $G$ is inferred by $\cA_{ti}$ to be a
  terminating mode for $P$.

\medskip

\noindent\textbf{$\Leftarrow$}~
Let $G$ be a terminating mode inferred by $\cA_{ti}$, we show that $\cA_{tc}$
proves termination of $G$. For this we show that for any $p(\bar x) \leftarrow
\varphi \in \posgd$ and $\beta\equiv p(\bar x) \leftarrow \pi,p(\bar y)
\in\sizebin$ there exists a decreasing set of arguments which is also
instantiated enough:
From the correctness of backwards analysis we know that $\varphi \models
\mu(p(\bar x)) \models \mu(\beta)$, and since
$\mu(\beta)$ was constructed in order to guarantee that at least one decreasing
arguments set for $\beta$ is instantiated enough, so there exists
$I'\in\dinf(\beta)$ such that $I' \subseteq I_{\mu(\beta)} \subseteq
I_{\varphi}$.
Since $\dinf(\beta)$ can be assumed without loss of generality to be extensive
$I_{\varphi}\in\dinf(\beta)$ and according to the accuracy requirements
$\dchk(I_{\varphi},\beta)$ answers ``yes''. So the termination condition holds
and $\cA_{tc}$ proves termination for $G$.
\end{proof}
\end{theorem}

In the case of our analysers, using the fast versions of $\dchk$ and $\dinf$,
checking is always as precise as inference. This follows as a simple result
from the definitions of $\dchk$ and $\dinf$. However, inference may be weaker
than checking.  The benchmark program \verb+rev_interleave+ in Table
\ref{table:t1} demonstrates this case.
Enhancing our analysers with \textsf{SVG} and \textsf{Extended SVG} for $\dchk$
and $\dinf$ respectively, would result in analysers which infer and check the
same sets of modes. This because both \textsf{SVG} and \textsf{Extended SVG}
are complete and hence of the same accuracy.
Note that we cannot make such a comparison for termination inference as
implemented in cTI because it is based on a different technique for inferring
termination conditions. While this technique seems equivalent to backwards
analysis, to make a formal comparison we would need to prove that it supports
the two claims \small\textbf{(BA$_1$)} and \small\textbf{(BA$_2$)}.

\section{Experimental Results}
\label{sec:exper}

This section describes an evaluation comparing our termination inference and
termination checking analysers. We also compare our analyser for termination
inference with the cTI \cite{Mesnard:sas01} analyzer.
For the experiments described, our analyser runs SICStus 3.7.1 on a Pentium III
500MHZ machine with 128MB RAM under Linux RedHat 7.1 (kernel 2.4.2-2). The cTI
analyser runs SICStus 3.8.4 on an Athlon 750MHz machine with 256MB RAM. The
timings for cTI are taken from \cite{Mesnard:sas01}.

Table \ref{table:t1} indicates analysis times in seconds for three blocks of
programs. The first two blocks correspond respectively to the programs from
Tables 2 and 5 in \cite{Mesnard:sas01}.
The third block contains two programs included to make a point detailed below.
The analysis parameters are the same as those reported in \cite{Mesnard:sas01}
--- term-size norm with widening applied every third iteration, except for the
programs marked by a $\star$ for which the list-length norm is applied and
widening is performed every fourth iteration.
The columns in the table indicate the cost for:

\begin{itemize}
  
\item[-]\textbf{Joint:} The activities common to termination checking and
  inference: preprocessing (reading, abstraction, computing sccs, printing
  results), size analysis (to approximate binary clauses) and groundness
  analysis (to approximate answers).
  Note that in TerminWeb, the checking component uses groundness analysis as
  described in \cite{Codish:JLP95} while the inference component uses a faster
  BDD based analyser.
  For the sake of comparison we consider the timing of the BDD based analyser
  for both checking and inference.
  
\item[-] \textbf{Inf:} The activities specific to termination inference:
  computing initial instantiation assertions as specified in
  Definition~\ref{def:assertion} (about 90\%) and performing backwards analysis
  (about 10\%).
  
\item[-] \textbf{Check:} The additional activities specific to termination
  checking for a single one of the top-level modes inferred to terminate.
  
\item[-]\textbf{Total Inf:} The total analysis time for inference using our
  analyser ($\textbf{Joint}+\textbf{Inf}$).

\item[-]\textbf{cTI:} The total analysis time for inference using cTI
  (timings as reported in \cite{Mesnard:sas01}).
\end{itemize}

\begin{table}[tbhp]
    \centering
\begin{tabular}{|l|lll|l||l|}
\hline
\textbf{Program}        & \textbf{Joint} & \textbf{Inf} & \textbf{Check} & \textbf{Total Inf} & \textbf{cTI}\\
\hline
permute                 &       0.13  & 0.01    & 0.04 &     0.14  & 0.15\\
duplicate               &       0.03  & 0.00    & 0.02 &     0.03  & 0.05\\
sum1                    &       0.05  & 0.01    & 0.02 &     0.06  & 0.18\\
merge                   &       0.19  & 0.02    & 0.04 &     0.21  & 0.26\\
dis-con                 &       0.09  & 0.01    & 0.04 &     0.10  & 0.24\\
reverse                 &       0.07  & 0.01    & 0.02 &     0.08  & 0.08\\
append                  &       0.06  & 0.00    & 0.00 &     0.06  & 0.09\\
list                    &       0.03  & 0.00    & 0.00 &     0.03  & 0.01\\
fold                    &       0.05  & 0.01    & 0.02 &     0.06  & 0.10\\
lte                     &       0.07  & 0.00    & 0.02 &     0.07  & 0.13\\
map                     &       0.05  & 0.00    & 0.02 &     0.05  & 0.09\\
member                  &       0.05  & 0.00    & 0.00 &     0.05  & 0.03\\
mergesort               &       0.44  & 0.02    & 0.06 &     0.46  & 0.43\\
mergesort$\star$        &       1.00  & 0.02    & 0.10 &     1.02  & 0.57\\
mergesort\_ap           &       0.63  & 0.04    & 0.30 &     0.67  & 0.79\\
mergesort\_ap$\star$    &       1.32  & 0.03    & 0.30 &     1.35  & 0.92\\
naive\_rev              &       0.10  & 0.00    & 0.02 &     0.10  & 0.12\\
ordered                 &       0.03  & 0.00    & 0.00 &     0.03  & 0.04\\
overlap                 &       0.06  & 0.00    & 0.02 &     0.06  & 0.05\\
permutation             &       0.12  & 0.01    & 0.04 &     0.13  & 0.15\\
quicksort               &       0.39  & 0.04    & 0.12 &     0.43  & 0.39\\
select                  &       0.10  & 0.00    & 0.01 &     0.10  & 0.08\\
subset                  &       0.11  & 0.00    & 0.02 &     0.11  & 0.09\\
sum2                    &       0.08  & 0.01    & 0.02 &     0.09  & 0.12\\
\hline
\hline
ann                     &       4.69  & 0.33    & 0.60 &     5.02  & 5.01\\
bid                     &       0.68  & 0.06    & 0.18 &     0.74  & 0.79\\
boyer                   &       2.70  & 0.05    & 0.14 &     2.75  & 3.53\\
browse                  &       1.01  & 0.15    & 0.37 &     1.16  & 1.81\\
credit                  &       0.49  & 0.05    & 0.15 &     0.54  & 0.61\\
peephole                &       4.59  & 0.09    & 0.58 &     4.68  &12.08\\
plan                    &       1.08  & 0.04    & 0.20 &     1.12  & 0.71\\
qplan                   &       11.04 & 0.54    & 3.43 &     11.58 & 7.30\\
rdtok$\oplus$           &       2.93  & 0.17    & 0.40 &     3.10  & 2.92\\
read$\ominus$           &       4.55  & 0.07    & 0.17 &     4.62  & 6.87\\
warplan$\oplus$         &       2.66  & 0.17    & 0.26 &     2.83  & 3.18\\
\hline
\hline
loop$\odot$             &       0.04  & 0.00    & 0.03 &     0.04  & -\\
rev\_interleave$\odot\otimes$&  0.21  & 0.02    & 0.03 &     0.23  & -\\
\hline
\end{tabular}

    \caption{Experimental Results}
  \label{table:t1}
\end{table}

\paragraph*{\textbf{Regarding precision}} 
For the first block of programs we infer exactly the same termination
conditions as cTI. For the second block (of larger programs), we infer the same
number of terminating predicates as does cTI, except for the last three
programs where a
``$\oplus$'' indicates that we infer termination for more predicates than does
cTI and a ``$\ominus$'' vice-versa. These differences stem from the fact that
the two analysers are based on slightly different components for approximating
loops. For all programs, in the first two blocks, our termination checker
verifies termination for the same set of modes as our termination inference
infers.
Note that for the second block of programs we count only the number of
terminating predicates in order to be consistent with the experiments reported
for cTI in \cite{Mesnard:sas01}.
The two programs in the third block demonstrate how the precision of the
$\dchk$ and $\dinf$ affect the precision of the analysis.
Here $\odot$ indicates that inference with \textsf{Extended SVG} is more
precise than inference with our simplified $\dinf$ procedure, and $\otimes$
indicates that our termination checking gives a more precise result than our
termination inference -- this is due to the fact that our choice of $\dchk$ and
$\dinf$ are not complete (as described in Section \ref{cvsi}).

\paragraph*{\textbf{Regarding timings}} 
The comparison of the columns \textbf{Total Inf} and \textbf{cTI} indicate that
TerminWeb and cTI are comparable for termination inference.
We note that the published results for cTI are obtained on a different machine,
the two analyzers are implemented using different versions of Sicstus Prolog
and they use different libraries for manipulating constraints.
For arithmetic constraints, TerminWeb uses the \textsf{clp(R)} library while
cTI uses the \textsf{clp(Q)} library. The prior is more efficient but may loose
precision. For Boolean constraints, TerminWeb uses the \textsf{BDD} library
described in \cite{Schachte:phd}, while cTI uses the Sicstus \textsf{clp(B)}
library. The prior is considerably faster.
More interesting is to notice the comparison of columns \textbf{Inf} and
\textbf{Check} which indicates that the cost of inferring all terminating modes
at once (computing assertions and apply backards analysis) is typically faster
than performing a termination check for a single mode.

\section{Related Work}
\label{sec:rw}

This paper draws on results from two areas: termination (checking) analysis and
backwards analysis. It shows how to combine components implementing these so as
to obtain an analyser for termination inference.
Termination checking for logic programs has been studied extensively (see for
example the survey \cite{DeSchreye:JLP94}).
Backwards reasoning for imperative programs dates back to the early days of
static analysis and has been applied extensively in functional programming.
Applications of backwards analysis in the context of logic programming are few.
For details concerning other applications of backwards analysis, see
\cite{king}.
The only other work on termination inference that we are aware of is that of
Mesnard and coauthors. The implementation of Mesnard's cTI analyser is
described in \cite{Mesnard:sas01} and its formal justification is given in
\cite{Mesnard:tocl}.

The two techniques (cTI and ours) appear to be equivalent. The real difference
is in the approach. Our analyser combines termination checking and backwards
analysis to perform termination inference.  This is a ``black-box'' approach
which simplifies design, implementation and formal justification. The
implementation reuses the TerminWeb code and an implementation of the backwards
analysis algorithm described and formally justified in \cite{king}.

Both systems compute the greatest fixed point of a system of recursive
equations. In our case the implementation is based on a simple meta-interpreter
written in Prolog. In cTI, the implementation is based on a $\mu$-calculus
interpreter.
In our case this system of equations is set up as an instance of backwards
analysis hence providing a clear motivation and justification
\cite{Mesnard:tocl}.

\section{Conclusion}
\label{sec:conc}

We have demonstrated that backwards analysis provides a useful link relating
termination checking and termination inference. This leads to a better
understanding of termination inference and simplifies the formal justification
and the implementation of termination inference.
We demonstrate this by enhancing the analyser for termination checking
described in \cite{CT:JLP} to perform also termination inference. We also
identify a simple condition which guarantees that termination inference can
infer all provably terminating modes when the corresponding analysers make use
of the same underlying analyses for size relations and instantiation
dependencies.

\paragraph*{\bf Acknowledgement} We thank Andy King, Fred Mesnard and
Cohavit Taboch for the useful discussions, as well as the exchange of
code and benchmarks.

%\bibliography{termin-inf.bib}
%\bibliographystyle{acmtrans}

\end{document}